# High-throughput calculations combining machine learning to investigate the corrosion properties of binary Mg alloys

Yaowei Wang[1,a], Tian Xie[2,a], Qingli Tang[1], Mingxu Wang[2], Tao Ying[2], Hong Zhu[1,2,*], Xiaoqin Zeng[2,*]

[1]*University of Michigan - Shanghai Jiao Tong University Joint Institute, Shanghai Jiao Tong University, 200240, Shanghai, RRC*

[2]*State Key Laboratory of Metal Matrix Composites, Shanghai Jiao Tong University, Shanghai, 200240, China, RRC*

# Abstract

Magnesium (Mg) alloys have shown great prospects as both structural and biomedical materials, while poor corrosion resistance limits their further application. In this work, to avoid the time-consuming and laborious experiment trial, a high-throughput computational strategy based on first-principles calculations is designed for screening corrosion-resistant binary Mg alloy with intermetallics, from both the thermodynamic and kinetic perspectives. The stable binary Mg intermetallics with low equilibrium potential difference with respect to the Mg matrix are firstly identified. Then, the hydrogen adsorption energies on the surfaces of these Mg intermetallics are calculated, and the corrosion exchange current density is further calculated by a hydrogen evolution reaction (HER) kinetic model. Several intermetallics, e.g. $Y_3Mg$, $Y_2Mg$ and $La_5Mg$, are identified to be promising intermetallics which might effectively hinder the cathodic HER. Furthermore, machine learning (ML) models are developed to predict Mg intermetallics with proper hydrogen adsorption energy employing work function ($W_f$) and weighted first ionization energy (WFIE). The generalization of the ML models is tested on five new binary Mg intermetallics with the average root mean square error (RMSE) of 0.11 eV. This study not only predicts some promising binary Mg intermetallics which may suppress the galvanic corrosion, but also provides a high-throughput screening strategy and ML models for the design of corrosion-resistant alloy, which can be extended to ternary Mg alloys or other alloy systems.

# Highlights

- A High-throughput screening workflow to investigate the corrosion properties of binary Mg alloys is developed.
- Several intermetallics, e.g. $Y_3Mg$, $Y_2Mg$ and $La_5Mg$, are identified to be promising intermetallics which might effectively hinder the cathodic HER.
- Machine learning models are applied to predict the hydrogen adsorption energy of Mg intermetallics, which can accelerate the high-throughput screening process.

∗ Corresponding authors.

E-mail address: hong.zhu@sjtu.edu.cn (H. Zhu), xqzeng@sjtu.edu.cn (X.Q. Zeng).

[a] Co-first authors, these authors contributed equally to this work.

# Graphical Abstract

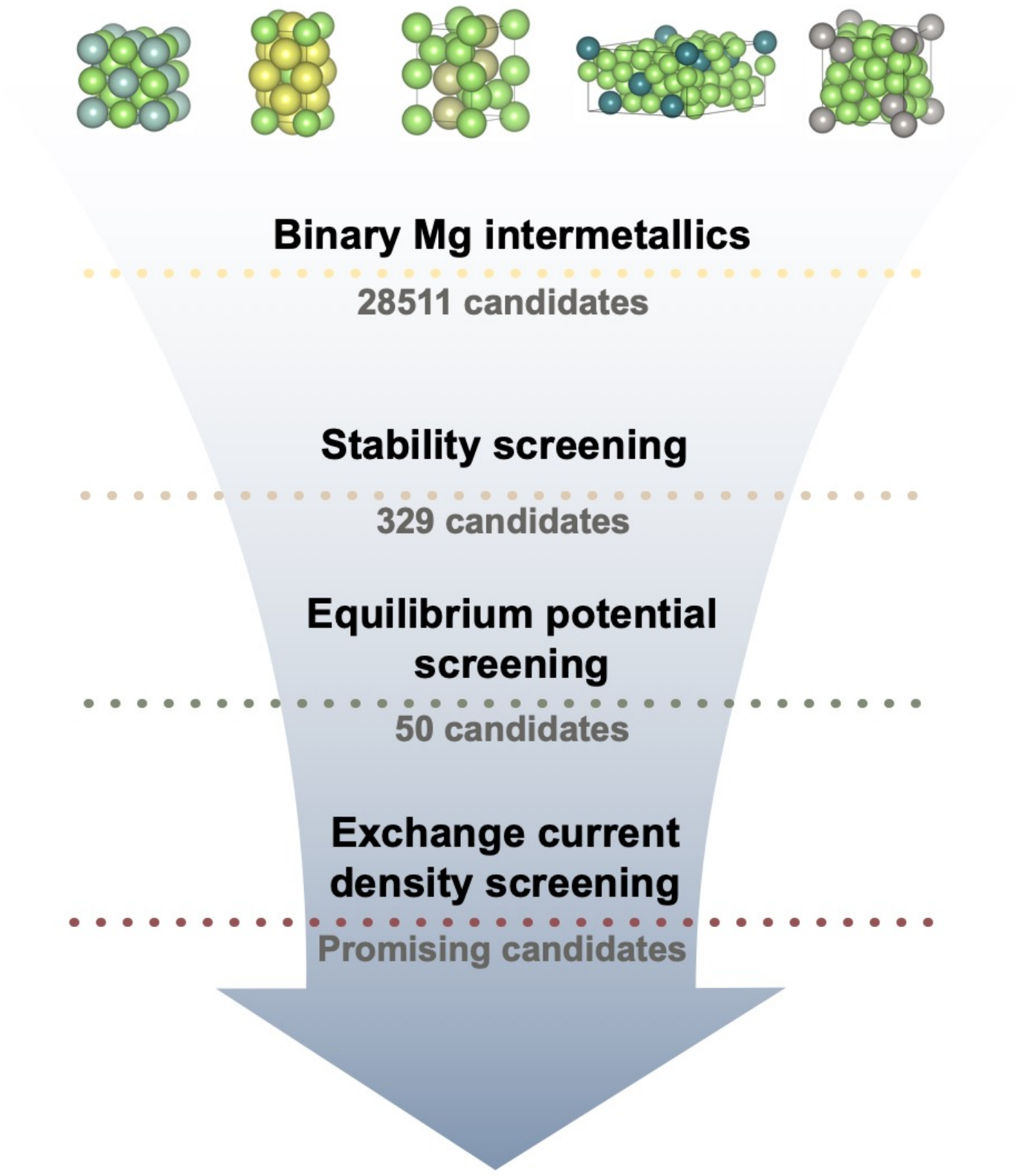

# 1. Introduction

As the lightest engineering structural materials, magnesium (Mg) alloys are considered as potential candidates in aerospace, automotive, electronics and biomedical fields[1–5]. However, continuous efforts have been made to improve the corrosion resistance of Mg alloys in service[6–9]. The poor corrosion resistance of Mg can be attributed to: (i) the high chemical activity, which provides a strong thermodynamic driving force for corrosion; and (ii) the incompact Mg oxide/hydroxide passivation layer, which cannot effectively protect the Mg matrix. As a consequence, galvanic couples can be easily formed due to uneven distribution of compositions, microstructure and crystal orientations in Mg alloys. Specifically, the intermetallic phases play an important role in the corrosion process and are deemed to accelerate the galvanic corrosion[10–12]. The galvanic corrosion of Mg alloys proceeds via the anodic dissolution reaction,

$$\mathrm{Mg} - 2\mathrm{e}^- \rightarrow \mathrm{Mg}^{2+} \tag{1}$$

and the cathodic hydrogen evolution reaction (HER)[13,14],

$$2\mathrm{H_2O} + 2\mathrm{e}^- \rightarrow 2\mathrm{OH}^- + \mathrm{H_2(g)} \tag{2}$$

Thus, the corrosion of Mg alloys can be mitigated by suppressing the anodic/cathodic reaction or reducing the thermodynamic driving force of the galvanic reaction. A possible strategy to slow down the cathodic reaction and hence the overall corrosion is through reducing the rate of HER. HER can proceed via either the Volmer-Tafel (Eq. (3) and (4)) or the Volmer-Heyrovsky mechanism (Eq. (3) and Eq. (5)). The corresponding reactions are

$$\mathrm{H}^+ + \mathrm{e}^- \rightarrow \mathrm{H}^* \quad \text{(Volmer reaction)} \tag{3}$$

$$2\mathrm{H}^* \rightarrow \mathrm{H_2(g)} \quad \text{(Tafel reaction)} \tag{4}$$

$$\mathrm{H}^* + \mathrm{H}^+ + \mathrm{e}^- \rightarrow \mathrm{H_2(g)} \quad \text{(Heyrovsky reaction)} \tag{5}$$

where H* indicates the adsorbed H atom on the cathode surface. Sabatier principle has revealed that the maximum HER rate can be obtained when there is a moderate binding energy between reactant and substrate[15]. This theory is further validated by the relationship between the exchange current density of HER and the metal-hydrogen bond strength[16]. Up to now, Sabatier principle has been widely applied in the field of catalyst screening[17,18]. Specifically, Nørskov et al. proposed that the free energy for hydrogen adsorption ($\Delta G_{H^*}$) could be used to predict the HER rate for a material surface and established the so-called volcano curve and corresponding kinetic model[19]. As for corrosion, to reduce the rate of cathodic HER, the $\Delta G_{H^*}$ should locate away from the summit of the volcano, indicating a relatively strong or weak H* adsorption to suppress the Volmer-Tafel or Volmer-Heyrovsky reaction[20].

It has been reported that the addition of alloying elements into Mg usually accelerates the HER reaction and hence the overall corrosion rate[6,21]. However, some elements such as arsenic (As) and germanium (Ge) have been recently reported to be corrosion inhibitors for Mg alloys[22–25]. Eaves et al. proposed As to be effective corrosion inhibitors for commercial Mg in sodium chloride electrolyte via obstructing hydrogen evolution[22]. Later, Birbilis et al. also observed similar phenomenon that As can reduce kinetics of the HER upon Mg [23]. Liu et al. investigated the role of Ge in binary and ternary Mg alloys and found that Ge could suppress the cathodic HER[24,25]. Recently, first-principles calculations have been applied to investigate the cathodic reactions on Mg and its alloys. Williams et al. explored thermodynamic barriers of HER on pure Mg and found the hydrogen recombination is the rate-determining step[26]. A continuous study by examining the thermodynamics of the water dissociation on dilutely alloyed Mg showed that the addition of Ge can make the water dissociation reaction endothermic, and hence may reduce the corrosion rate[27]. Sumer et al. and Yuwono et al. compared HER reaction barrier of some common binary Mg alloy systems and reported that As and Ge can suppress cathodic kinetics [28,29]. In spite of these limited success[30,31], a more comprehensive study about the influence of intermetallics, which is a common existing form for elements in Mg alloys, on HER of Mg alloys is still lacking.

High-throughput calculations have become an effective tool to screen promising candidates in discovery of medicines, catalysts and battery electrolytes[32–34]. Montoya et al. developed a high-throughput workflow for the adsorption energy calculation, such method could accelerate the discovery of new high-efficient catalysts[35]. As for corrosion, Qi et al. performed high-throughput calculations to search for elements in Fe impurity phases which can inhibit the cathodic HER of Mg alloys[20]. Although the surfaces with different terminations and adsorption sites can be simulated in a high-throughput way[35,36], these DFT-based surface calculations are still time-consuming, especially considering a large pool of intermetallic compounds. Take calculation the hydrogen adsorption energies of $MgZn_2$ as an example [30], up to Miller indices (111), totally 23 surfaces with different terminations should be considered and average 6 distinct H adsorption sites on each surface should be calculated. This yields around 120 DFT calculations for $MgZn_2$. To address this challenge, data-driven methods such as machine learning (ML) could be adopted to accelerate the screening process. For example, Raccuglia et al. built ML model with chemical information contained in historical reactions and accurately predicted if the reaction can proceed [37]. Zahrt et al. trained ML model and predicted the higher-performing catalysts, which is potential to change the selection way of catalysts [38]. Therefore, the application of ML is prospective in predicting Mg alloy systems with strong corrosion resistant behavior.

In this work, a three-step screening is applied to search for promising binary Mg intermetallics, the presence of which could have a small thermodynamic driving force and low HER kinetic for the galvanic corrosion of Mg alloys. Potential Mg intermetallic phases compounds obtained from open databases (Materials Project[39], OQMD[40] and AFLOW[41]) have been screened based on three criteria, namely the phase stability screening, the equilibrium potential screening and the HER kinetics screening. To further accelerate screening process, ML algorithms are applied to establish the correlation between the H adsorption energy with physical and chemical properties such as work function and weighted first ionization energy. It is expected that this computational screening based on DFT calculations and ML predictions can effectively guide the design of corrosion-resistant binary Mg alloy and inspire the design of other corrosion-resistant alloy systems.

# 2. Methods

## *2.1. Criteria to screen binary Mg intermetallics inhibiting Mg corrosion*

The following three criteria were applied to screen binary Mg intermetallics which could inhibit Mg corrosion. Firstly, the intermetallics should be thermodynamically stable or metastable to ensure the probability of experimental synthesis. In this study, thermodynamic stability was evaluated by energy above the convex hull ($E_{Hull}$)[42,43]. Convex hull is a plot of formation energy with respect to the composition which connects phases with lowest formation energy than other phases. Phases lying on the convex hull are thermodynamically stable ($E_{Hull}$ equals to 0) and the ones above the convex hull are either metastable or unstable[44]. Secondly, the equilibrium potential difference between intermetallic phases and Mg matrix should be small to minimize the driving force of galvanic corrosion. Thirdly, the exchange current density calculated by the kinetic model proposed by Nørskov should be small enough to slow down the cathodic HER[19,45]. For the third criterion, the surface energies of low index surfaces (Miller indices up to (111)) of intermetallics were calculated and the most stable surface of each intermetallic was retained. Then, the distinct H adsorption sites on the stable surfaces were enumerated by pymatgen[35] codes and the lowest adsorption energy was adopted to calculate the exchange current density.

## *2.2. Frist-principles calculations*

In this work, all DFT calculations were carried out by utilizing projector augmented wave (PAW)[46] method implemented in the Vienna *ab initio* simulation package (VASP)[47]. The exchange-correlation functional is described by generalized gradient approximation (GGA)[48] with Perdew-Burke-Ernzerhof (PBE) approach[49]. When operating high-throughput computation, the cut-off energy of plane wave was set at 480 eV. Considering different lattice structures of intermetallics, Gamma-centered k-point grids were automatically generated by pymatgen codes[36]. The convergence criteria of energy and force are set to $10^{-4}$ eV and 0.02 eV/Å, respectively. Atoms 3 Å away from the surface were fixed during the structural optimization. The details of convergence tests can be found in Appendix Fig. A1 and A2. To avoid the interaction caused by periodic images, 15 Å vacuum layer was employed along Z direction. The calculation method of the surface energy can be found in our previous works[30,31]. The hydrogen adsorption energy ($E_{ads}$) was calculated by

$$E_{ads} = E_{slab*H} - E_{slab} - \frac{1}{2} E_{H_2} \tag{6}$$

where $E_{slab*H}$, $E_{slab}$ and $E_{H_2}$ are the DFT energies of the slab with one hydrogen adatom, the bare slab and the hydrogen molecule, respectively. The free energy for hydrogen adsorption can be calculated by

$$\Delta G_{H^*} = E_{ads} + \Delta E_{ZPE} - T\Delta S_H \tag{7}$$

where $\Delta E_{ZPE}$ and $\Delta S_H$ are the difference in zero-point energy and entropy between the adsorbed and the gas phase, respectively. $\Delta E_{ZPE} - T\Delta S_H$ is calculated to be 0.19 eV for most of H adsorption on Mg or Mg common intermetallics. This value is taken to be representative for all the binary intermetallics studied here, which means $\Delta G_{H^*} = E_{ads} + 0.19$ eV. Detail information can be found in Appendix table A2.

## 2.3. Machine learning

Support vector regression (SVR) and k nearest neighbors (KNN) ML algorithm were adopted to predict $E_{ads}$ on Mg intermetallics surface[50,51]. All the input data were normalized by z-score method to decrease the influence of data distribution range (electronegativities vary from 1.11 to 2.0 whereas relative molecular masses of intermetallics span from 113 to 1458). Coefficient of determination ($R^2$) and root mean square error (RMSE) were applied to test the stability and accuracy of the model. To prevent over-fitting, the dataset was randomly separated into 90% training data to train the model and 10% test data to evaluate the model performance. Moreover, 500 times random divisions of training and test dataset were performed and the final performance were measured by the average of all results. The hyperparameters, which have a crucial effect on model performance, were carefully selected by grid search method and the details were listed in Table A1 of appendix. The definitions of data normalization, $R^2$ and RMSE were also supplied in Appendix. For ML implementation, we used an open source Python module, Scikit-learn[52].

# 3. Results and discussion

## *3.1. Thermodynamic stability screening*

The designed high-throughput workflow for screening binary Mg intermetallics which could inhibit galvanic corrosion is shown in Fig. 1. At the beginning of the screening process, all potential candidates are collected from three materials data repositories, namely Materials Project[39], OQMD[40] and AFLOW[41] and there are 28511 binary Mg intermetallics found in total. After removing duplicate entries and the structures with more than 30 atoms[53], 995 intermetallics are retained. To filter out unstable candidates, the threshold of $E_{Hull}$ to distinguish between stable and unstable intermetallics is taken as 50 meV/atom[54,55] and the reference energies of stable phases are employed from Materials Project database. For the intermetallics from OQMD and AFLOW, $E_{Hull}$ is recalculated using pymatgen codes to guarantee entries from different database comparable. The convergence parameters adopted in this work are consistent with those of the Materials Project. After phase stability screening based on $E_{Hull}$ smaller than 50 meV/atom, there are 329 binary Mg intermetallics left among 995 candidates, and in principle, they are all possible to be experimentally synthesized.

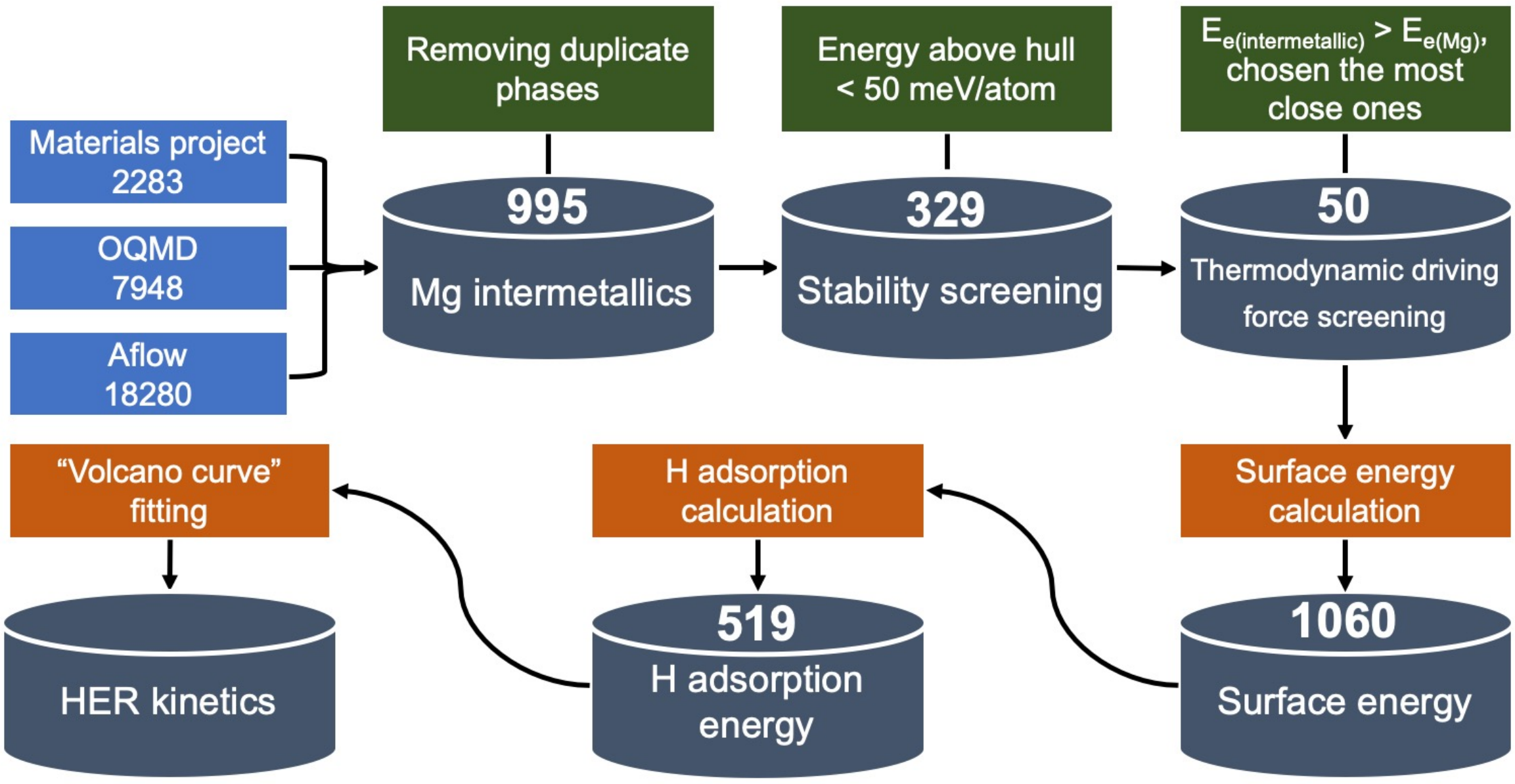

Fig. 1. The designed high-throughput workflow for screening binary Mg intermetallics which could inhibit galvanic corrosion. Total 28511 binary Mg intermetallics are collected from Materials Project[56], OQMD[40] and AFLOW[41] database. After removing duplicate entries and the structures with more than 30 atoms, 995 binary Mg intermetallics are retained. The first screening step is phase stability screening based on the energy above convex hull and 329 out of 995 intermetallics are kept. The second step is the equilibrium potential screening and 50 intermetallics serving as cathode during the galvanic corrosion with the smallest equilibrium potential difference

with respect to the Mg matrix are reserved. Lastly, the adsorption sites on most stable surfaces are enumerated by pymatgen codes and the most stable H adsorption energies are adopted to calculate exchange current density via the HER kinetic model proposed by Nørskov[19,45].

## *3.2. Equilibrium potential screening*

In commercial Mg alloys, most intermetallic phases are reported to be nobler than Mg matrix and serve as the local cathode during galvanic corrosion processes[57]. Sudholz et al. investigated the potentiodynamic polarization curves of binary Mg intermetallics, revealing that most intermetallics possess a higher corrosion potential than pure Mg except $Mg_2Ca$[10]. It is believed that a higher corrosion potential difference between the Mg matrix and the intermetallics contributes to a higher corrosion tendency[58]. Hence, the equilibrium potential difference between Mg and intermetallics is applied to evaluate the thermodynamic driving force of galvanic corrosion in this work. The dissolution reactions of Mg and intermetallics are assumed to happen in neutral solution, and at the room temperature with all ionic concentration of $10^{-6}$ mol/L. Fig. 2 shows the 50 out of 329 Mg intermetallics with the smallest equilibrium potential difference with respect to the Mg matrix. The results indicate that most of intermetallics containing rate-earth (RE) elements show lower corrosion tendency. It is noteworthy that as previous works reported, some elements which could "poison" cathodic HER, e.g. As and Ge[23,24], were not retained after the thermodynamic screening. We checked the equilibrium potential of some "poisonous" Mg intermetallics, e.g. $Mg_3As_2$, $MgAs_4$ and $Mg_2Ge$, and found that all of them serve as local cathode during galvanic corrosion with rather high potential differences with Mg matrix, and thus were excluded. A more comprehensive study including corrosion properties of all intermetallics will be conducted in future to capture those intermetallics with very low kinetic but relatively large thermodynamic driving force for galvanic corrosions. The details of equilibrium potential calculation could be found in our previous work[30] and the calculated equilibrium potential results of 329 Mg binary intermetallics are listed in Table A3.

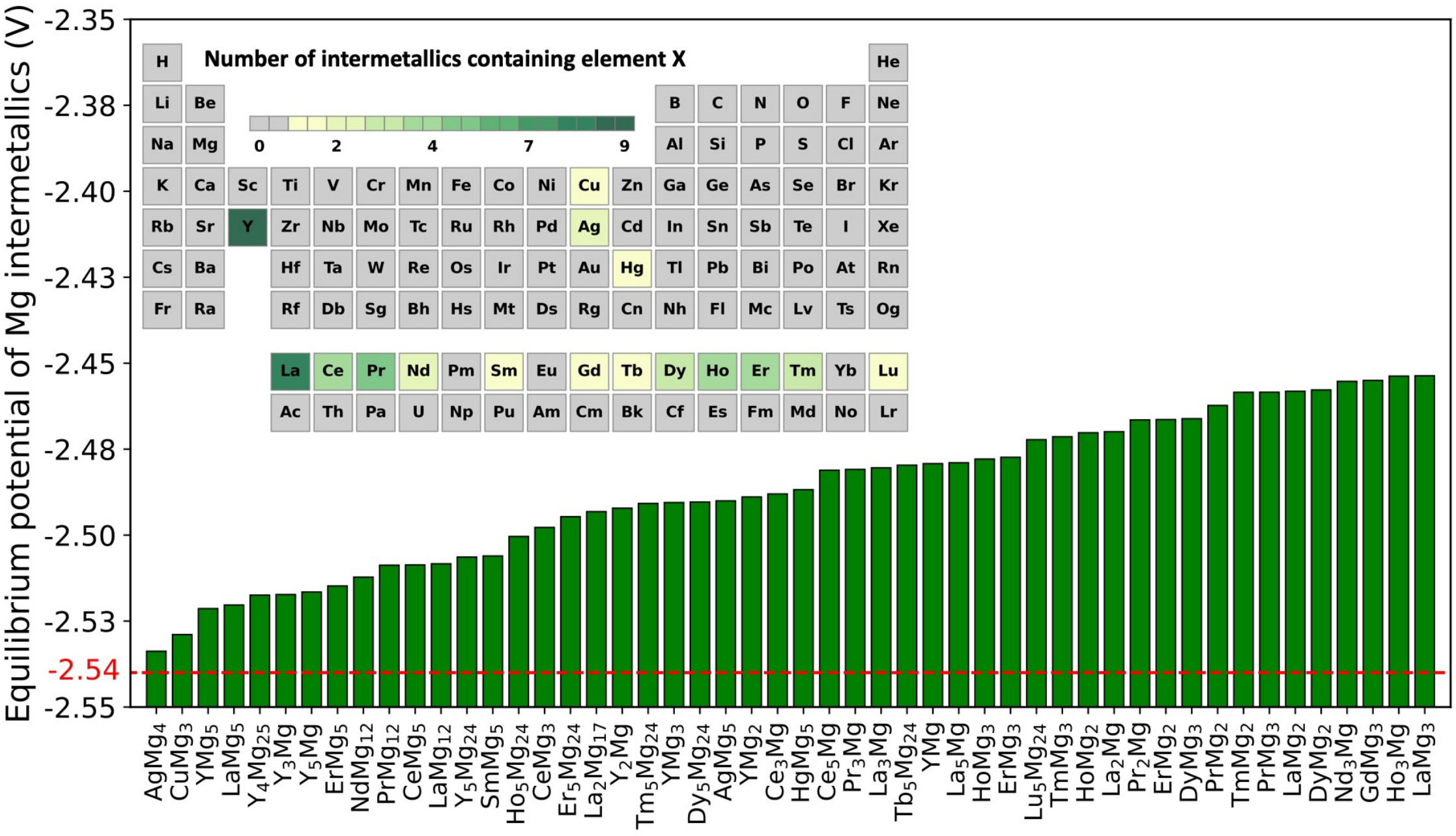

Fig. 2. Calculated equilibrium potential of 50 thermodynamically stable binary Mg intermetallics evaluated by energy above convex hull. The red dashed line represents the equilibrium potential of pure Mg in neutral solution and at room temperature. All screened intermetallics possess higher equilibrium potential of Mg matrix, which are deemed to serve as cathode during the galvanic corrosion. The inset shows the screened elements in the periodic table as well as the number of screened intermetallic compounds for a given binary alloy systems.

### *3.3. HER kinetics screening*

In actual applied environment, Mg usually serve as anode accompanied by the hydrogen evolution on nobler region, i.e. intermetallic phase, as the main cathodic reaction. Therefore, apart from the equilibrium potential difference, the reaction rate of cathodic HER is another important screening criterion for corrosion-resistant Mg alloy. To study the HER kinetics on the intermetallic surface, the most stable surface termination was identified for each intermetallic compound. All the low-index surfaces (Miller indices up to (111)) of 50 selected Mg intermetallics with different terminations were considered and 1060 surface energies were obtained accordingly. For most Mg intermetallics, the most stable surface is found to be (100) or (111) surface. Subsequently, 519 adsorption sites on the most stable surface of Mg intermetallics were enumerated by the pymatgen codes and the most stable adsorption configurations were obtained based on high-throughput simulations. The DFT-calculated surface energies and $E_{ads}$ could be found in Table A3.

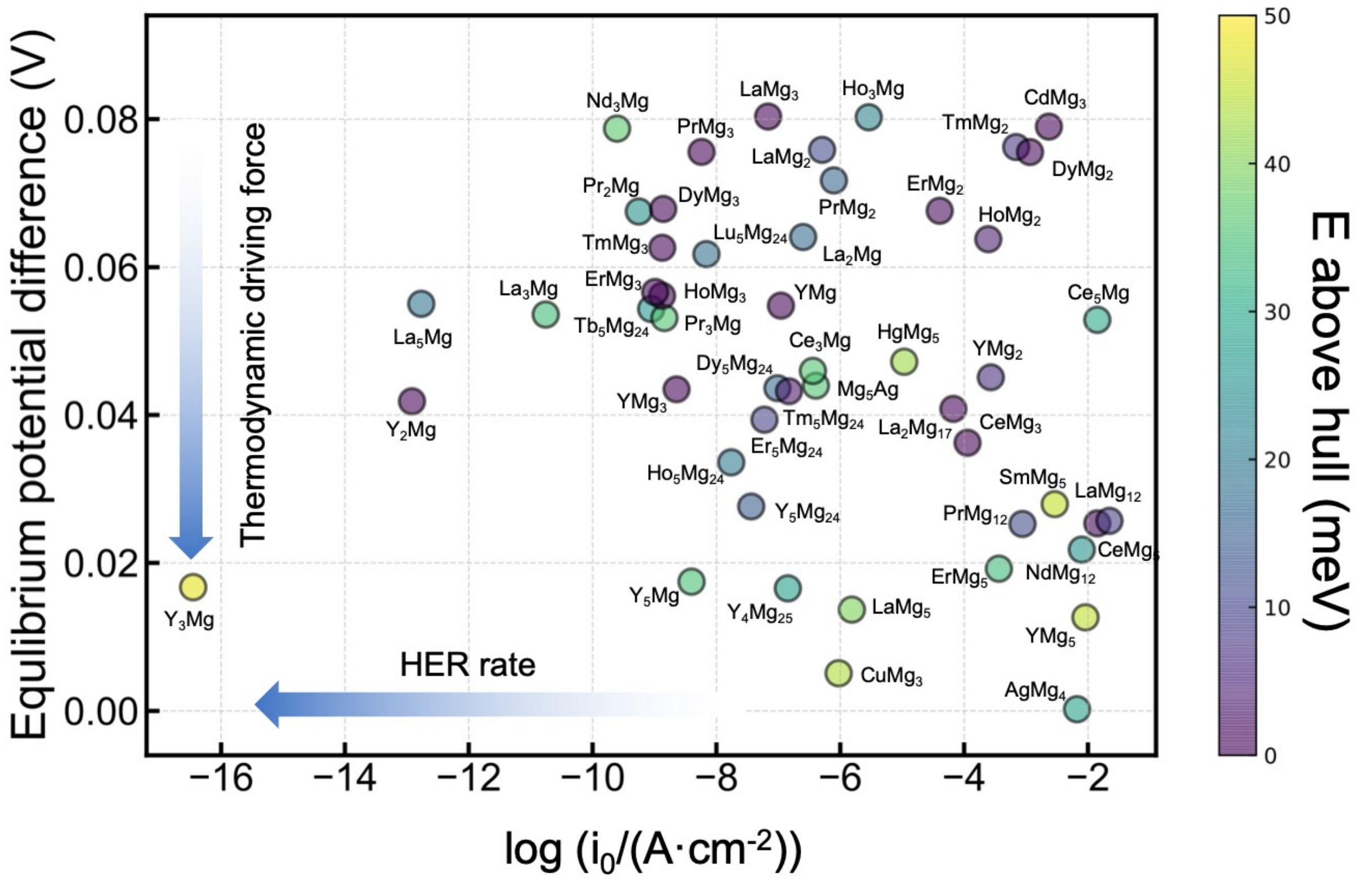

Fig. 3. Comprehensive screening results containing phase stability, equilibrium potential difference and calculated exchange current density. The color of dots represents the stability of intermetallics quantified by $E_{Hull}$. The purple and yellow color corresponds to stable ($E_{hull} = 0$ meV) and semi stable ($E_{hull} = 50$ meV) intermetallics, respectively. The HER exchange current densities of Mg intermetallics are calculated by the HER kinetic model proposed by Nørskov[19,45].

Based on the HER kinetic model proposed by Nørskov et al., the calculated $\Delta G_{H^*}$ are further used to calculate the exchange current density, which could be an indicator for HER rate on different intermetallics[19]. The comprehensive screening results containing phase stability, equilibrium potential difference and predicted current density are shown in Fig. 3. The color of dots represents the stability of intermetallics quantified by $E_{Hull}$, X-axis represents the exchange current density calculated by the HER kinetic model, and Y-axis represents the equilibrium potential difference between intermetallics and Mg matrix. In principle, smaller equilibrium potential difference (thermodynamic driving force control) and lower exchange current density of HER (kinetics control) will contribute to weaker galvanic corrosion. Therefore, the promising candidates lie in the lower left corner of Fig. 3, such as several Mg-Y and Mg-La intermetallics including $Y_3Mg$, $Y_2Mg$ and $La_5Mg$.

Several studies have reported the corrosion properties of Mg-rare earth (RE) alloys. Liu et al. demonstrated the corrosion resistance of Mg alloys would be improved by Lutetium (Lu) rare earth element, which formed $Lu_5Mg_{24}$ intermetallic phases[59]. To valid this kinetic model, we collected the polarization curves of intermetallics presented in common Mg alloy systems[10]. The exchange current density $i_0$ of Mg intermetallics are obtained by extrapolating the linear region of Tafel plots to the reversible potential of the HER at pH=7. Fig. A5 shows the calculated exchange current density of Mg intermetallics in this work is linearly correlated with the experimental exchange current density obtained from polarization curves, indicating this kinetic model can be a guide to estimate the corrosion cathodic reaction rate for Mg and Mg intermetallics.

### *3.4. Feature engineering*

Correlation analysis is applied to further investigate the factors that influence $E_{ads}$ of the Mg intermetallics. Fig. 4 shows the statistical distribution plots between $E_{ads}$ and some structural information, e.g. sum of squares of Miller indices and the crystal structure. Fig. 4a indicates that the adsorption of H tends to be unstable with the increasing of the sum of squares of Miller indices, which could be intuitively explained by bond-order conservation theory[60,61]. It is expected that low-index surfaces are less coordinated than high-index surfaces, and thus the associated interaction between surface atoms and adsorbates is stronger. In other words, the fewer dangling bonds a surface atom has, the less it will bind adsorbates. Additionally, as shown in Fig. 4b, the orthogonal lattices including BCC and FCC structures are more likely to possess the most negative or positive $E_{ads}$, which may be the promising crystal structures of intermetallics suppressing HER kinetics.

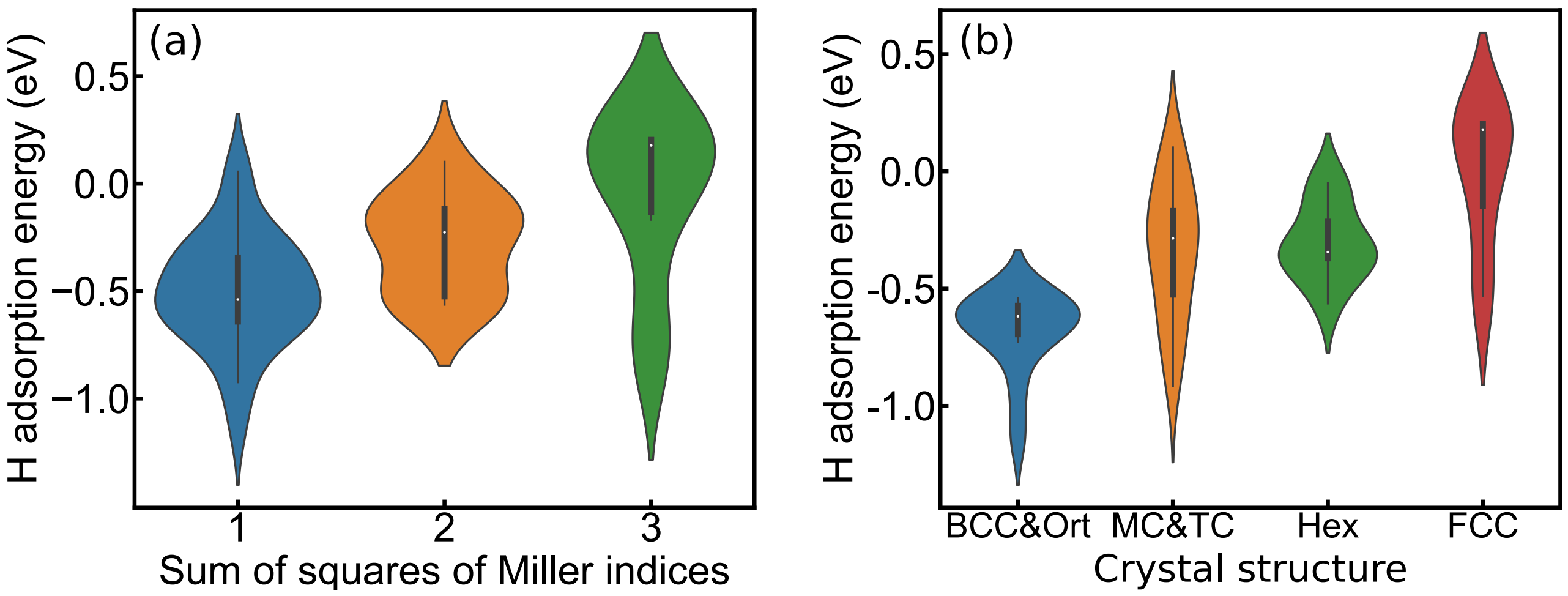

Fig. 4. H adsorption energies distribution of (a) sum of Miller indices squares and (b)crystal structure. The abbreviations BCC&Ort in (b) represents Body Centred Cubic and Orthogonal structure, MC&TC represents Monoclinic and Tetragonal structure, Hex represents Hexagonal structure and FCC represents Face Centred Cubic structure.

As aforementioned, $E_{ads}$ plays an important role in the cathodic reaction of galvanic corrosion. To further accelerate the screening process and save computational cost, 18 primary features which may be correlated with H adsorption were adopted to predict $E_{ads}$. These features can be roughly divided into two categories. The first type of features needs DFT calculation, such as surface work function ($W_f$) and *d*-band center (*d*). The second type of features are elemental properties, such as the electronegativity (EN) of X for Mg-X intermetallics. Moreover, the intermetallic electronic properties are also represented by Weighted electronegativity (WEN), Weighted electron affinity (WEA) and Weighted first ionization energy (WFIE), which are calculated by sum of pure elemental properties times corresponding elemental molar ratio. These features were selected as we think they are correlated with the surface bonding to some extent. For instance, $W_f$ and $\gamma$ are related to surface stability[62,63], *d*-band center has been proven particularly useful in understanding bond formation between adsorbates and transition metal substrates[64]. The features and their corresponding abbreviations are summarized in Table 1.

Table 1 DFT calculated and elemental features adopted in this work.

| DFT calculated features | | Elemental features | |
|---|---|---|---|
| $\Delta H_f$ | Formation enthalpy (eV/atom) | **EN** | Electronegativity |
| $E_g$ | Band gap (eV) | **EA** | Electron affinity (eV) |
| $\gamma$ | Surface energy (J/m$^2$) | **FIE** | First ionization energy (eV) |
| $E_{hull}$ | E above hull (eV/atom) | **AR** | Atomic radius (Å) |
| $W_f$ | Work function without H adatom (eV) | **AIR** | Average ionic radius (Å) |
| $d$ | d-band center without H adatom (eV) | **RAM** | Relative molecular mass |
| $S_m$ | Sum of squares of Miller indices | **WEN** | Weighted electronegativity |
| $E_e$ | Equilibrium potential | **WEA** | Weighted electron affinity (eV) |
| $B_H$ | Bader charge transfer to H adatom | **WFIE** | Weighted first ionization energy (eV) |

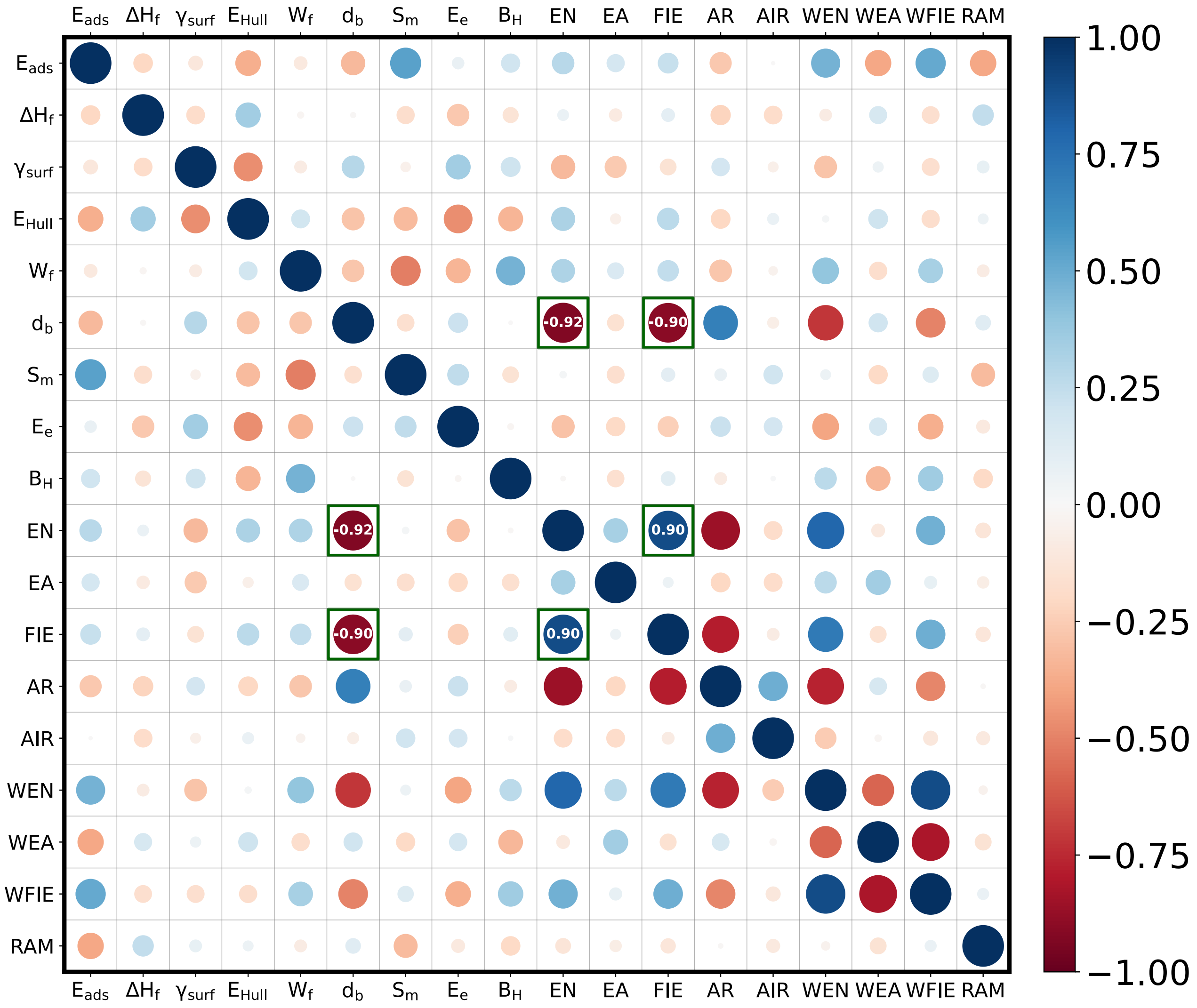

Fig. 5 Pearson correlation coefficient (PCC) correlation map of 18 features. The blue and red colors represent positive and negative correlations, respectively. Darker color and bigger circles indicate stronger correlation. The green square frames mean two features are highly correlated with each other ($|p| > 0.9$).

Feature selection plays a crucial role for training model with excellent performance[65], and each feature should be independent to represent the certain physical or chemical properties. Additionally, the feature with too much noise data should also be excluded to maintain the predictive capacity of the model. Following the data processing principles, the $E_g$ and $E_{hull}$, which include some zero values, are firstly removed. The correlations of pairwise features evaluated via Pearson correlation coefficient (PCC) are shown in Fig. 5. For the two features with the $|p|$ larger than 0.9, we only keep one in our feature set. Based on the above consideration, the number of features is decreased from 18 to 14. We aim at predicting $E_{ads}$ with simple features and low computational cost. Hence, all possible subsets of Ω dimensional features (Ω ranges from 1 to 5) were exhaustively enumerated to identify the feature subset giving rise to minimal prediction error.

### *3.5. Machine learning models training and generalization*

As shown in Fig. 6, SVR and KNN algorithms are employed to build regression models to predict $E_{ads}$ on binary Mg intermetallic surfaces. The cross-validation RMSE of KNN and SVR models based on different dimensional features are illustrated in Fig. 6a and Fig. 6d. Two ML algorithms possess comparable prediction capacity for $E_{ads}$ with the average RMSE of 0.13 eV when utilizing two features. According to Fig. 6b and Fig. 6e, with the increasing of utilized features, the initial increase in accuracy demonstrates more features would improve the prediction capacity of the model, while further increasing the number of features will decrease the prediction accuracy due to possible over-fitting. The best performance of the KNN and SVR models are given by 2 and 4 features, respectively. Interestingly, the best two features of KNN and SVR algorithms are both $W_f$ and WFIE, indicating they are highly correlated with $E_{ads}$. The details about best Ω dimensional features are shown in Fig. A1. Starting from the best two-feature subset, adopting additional features slightly improves the prediction capacity of models, but increases the complexity of the model at the same time. Consequently, the best two features for two ML algorithms, i.e. $W_f$ and WFIE, were chosen for building ML models without sacrificing much accuracy and generalizability.

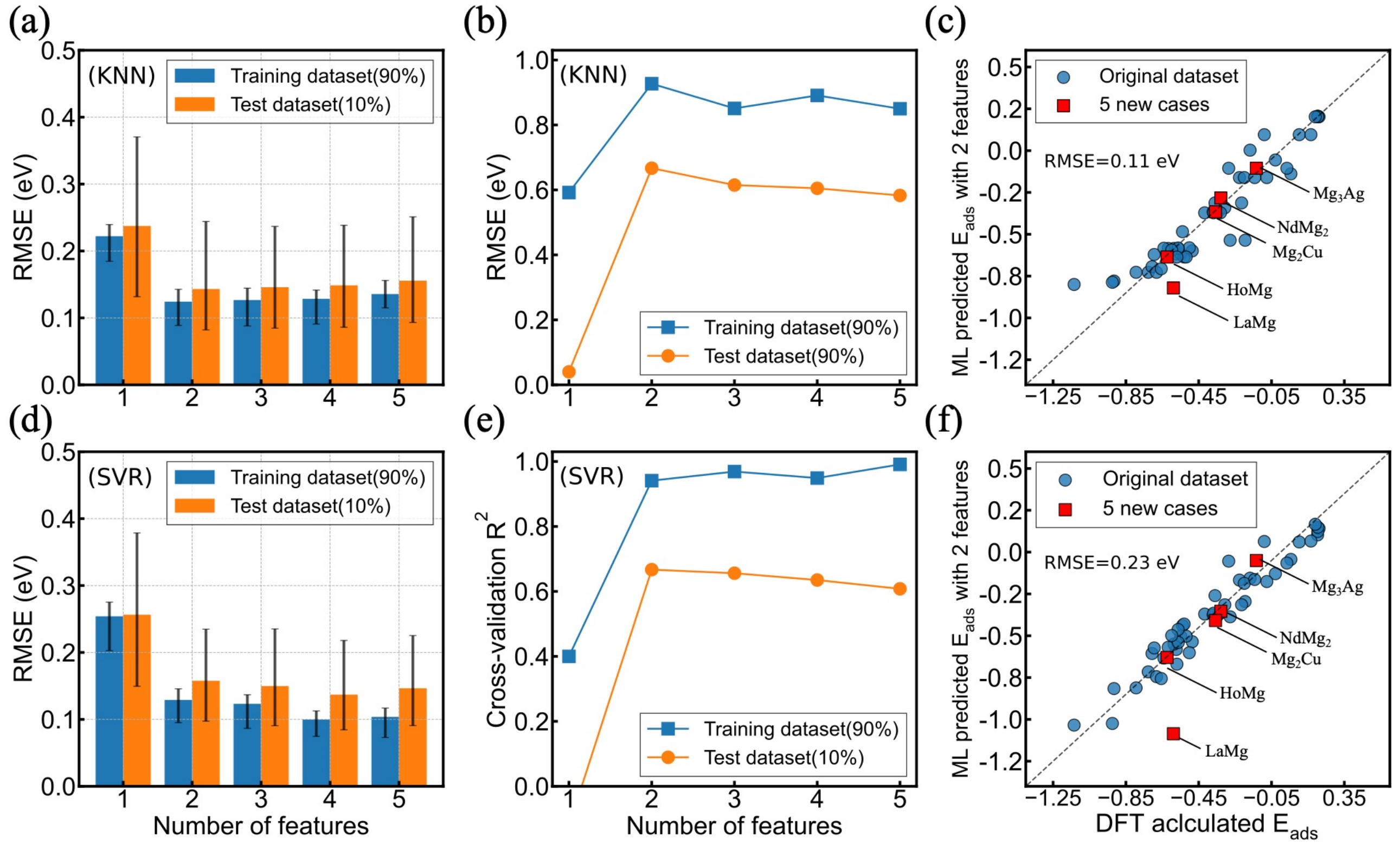

Fig. 6. Average RMSE of 500 times random divisions of training and test dataset for (a) KNN and (d) SVR model with Ω-dimensional features (Ω ranges from 1 to 5). Error bars indicate the maximum and minimum Error in 500 repetitions of training. The cross-validation $R^2$ of best performance model containing Ω-dimensional features for (b) KNN and (e) SVR models. Parity plots comparing DFT-computed $E_{ads}$ against ML-predicted $E_{ads}$ via (c) KNN and (f) SVR models, as well as 5 new intermetallics not included in the training dataset.

The above results indicate our ML models perform well on training data. To further test the generalization of models, $E_{ads}$ of the 5 new intermetallic compounds, LaMg, HoMg, $Mg_2Cu$, $NdMg_2$ and $Mg_3Ag$, which possess close equilibrium potential with Mg matrix, were calculated based on DFT simulations (see Table A4 for the results, along with corresponding work function and weighted first ionization energy). The comparison of the $E_{ads}$ from DFT calculations and ML predictions are shown in Fig. 6c and Fig. 6f, with average RMSE of 0.11 and 0.23 eV, respectively. Actually, $W_f$ and WFIE represent the energies needed to remove electrons from metal or neutral atoms in gas phase, which can be essentially understood as the ability of intermetallics to bind electrons. The $E_{ads}$, as well as HER kinetics, is largely depended on the binding strength between H atom and intermetallic surface. Previous studies also employed the $W_f$ or FIE as features to successfully predict adsorption energies of small molecules[66,67]. Takigawa et al. utilized extra tree regression algorithm and 12 features to predict H adsorption energy on doped Cu with the average testing RMSE of 0.17 eV[68]. As shown in Fig. 6c and Fig. 6f, the prediction error of these new compounds is comparable to that of the original test dataset, demonstrating our ML model is robust to predict H adsorption energy on new cases and thus instructive for screening the corrosion-resistant binary magnesium alloys with intermetallics.

# 4. Conclusion

By means of first-principles calculations, a three-step high-throughput screening, namely stability screening, equilibrium potential screening and HER kinetics screening, was performed to search for promising corrosion-resistant binary Mg alloys with intermetallics. We found that most Mg-RE binary intermetallics possess close equilibrium potential with Mg matrix, which will serve as a weak cathode. Specifically, several intermetallics, e.g. $Y_3Mg$, $Y_2Mg$ and $La_5Mg$, can hinder the galvanic corrosion reaction due to the relatively small thermodynamic driving force and low HER kinetics. Moreover, ML models have been applied to predict the hydrogen adsorption energy of Mg intermetallics, which can accelerate the high-throughput screening process. The robustness and generalization of ML models was tested on new binary Mg intermetallics. This work by combining

DFT, thermodynamics and kinetics analysis and ML not only predicts some promising binary Mg intermetallics which can hinder the galvanic corrosion, but also provide a high-throughput screening strategy for corrosion-resistant metal alloy design, which could be instructive for future experiments.

# Data availability

The codes and data generated in this work are available at https://github.com/ywwang0/High-throughput-screens-corrosion-resistant-binary-magnesium-alloy.

# Acknowledgements

The research was financially supported by the National Key Research and Development Program of China (No. 2016YFB0701202, No. 2017YFB0701500 and No. 2020YFB1505901), National Natural Science Foundation of China (General Program No. 51474149, 52072240), Shanghai Science and Technology Committee (No. 18511109300), and Science and Technology Commission of the CMC (2019JCJQZD27300). First-principles calculations were carried out with computational resources from Shanghai Jiao Tong University Super Computer Center. H. Zhu also thanks to the financial support from the University of Michigan and Shanghai Jiao Tong University joint funding, China (AE604401) and Science and Technology Commission of Shanghai Municipality (No. 18511109302).

# Declaration of interest

The authors declare that there are no competing interests.

# Appendix for "High-throughput computation combining machine learning to investigate the corrosion properties of binary Mg alloys"

Yaowei Wang[1a], Tian Xie[1b], Qingli Tang[a], Mingxu Wang[b], Tao Ying[b], Hong Zhu[a,*], Xiaoqin Zeng[b,*]

[a]University of Michigan - Shanghai Jiao Tong University Joint Institute, Shanghai Jiao Tong University, 200240, Shanghai, RRC

[b]National Engineering Research Center of Light Alloy Net Forming, Shanghai Jiao Tong University, 200240, Shanghai, PRC.

∗ Corresponding authors.

E-mail address: hong.zhu@sjtu.edu.cn (H. Zhu), xqzeng@sjtu.edu.cn (X.Q. Zeng).

# Data Normalization:

$$x_i' = \frac{x_i - \mu_x}{\sigma_x}$$

where $x_i'$ and $x_i$ are normalized and primary feature vector. $\mu_x$ and $\sigma_x$ are mean value and standard deviation of the feature vector.

# Formula of R2 and RMSE

Coefficient of determination:

$$R^2 = 1 - \frac{\sum_{i=1}^{n}(y_{true} - y_{pre})^2}{\sum_{i=1}^{n}(y_{true} - \bar{y}_{true})^2}$$

Root-mean-squared error:

$$RMSE = \sqrt{\frac{1}{n}\sum_{i=1}^{n}(y_{true} - y_{pre})^2}$$

Where n, $y_{true}$, $y_{pre}$ are the number of samples, DFT-calculated H adsorption energy, predicted H adsorption energy via machine learning, respectively.

## Convergence test

In high high-throughput calculations, the calculation parameters should be carefully selected to balance the computational accuracy and efficiency and a convergence tests have been conducted. Here we take MgSc intermetallic with cubic crystal system as an example to shown the results of convergence test. (Pearson symbol of MgSc is cP2 and its space group is $Pm\bar{3}m$)

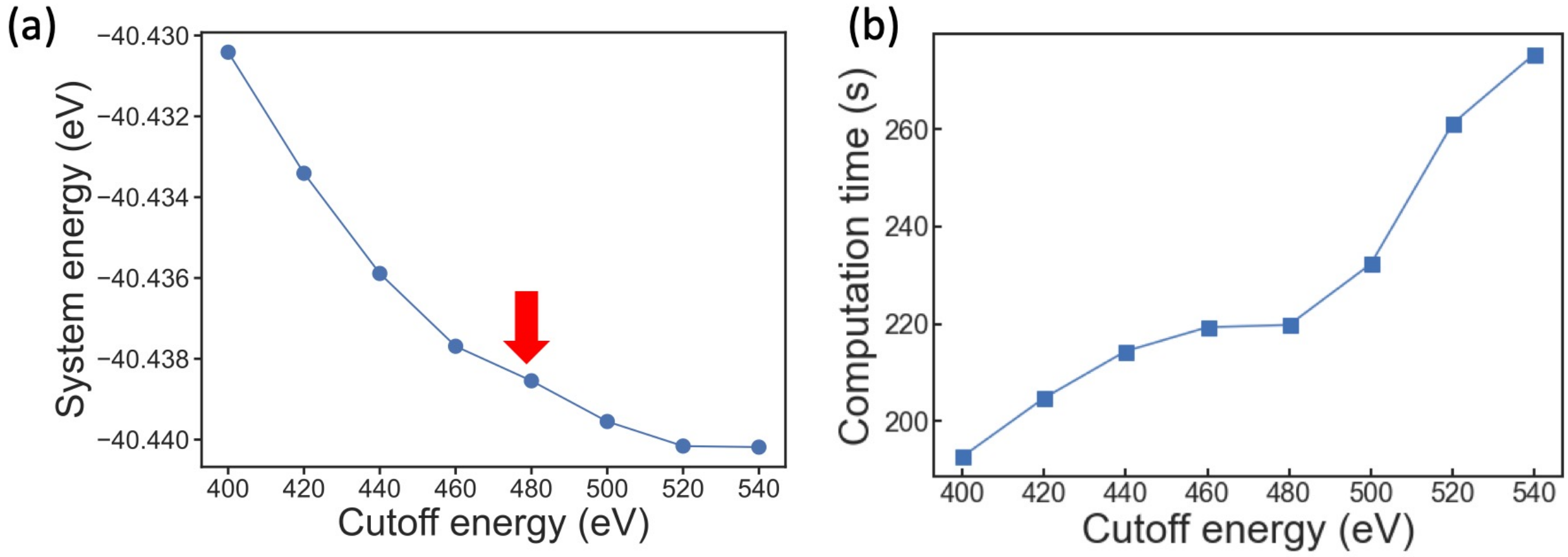

Figure A1. Convergence curves of MgSc slab model with respect to cutoff energy.

According to the VASP manual, the cut-off energy of plane wave should be set as around 1.3*ENMAX. In our calculation system, element Ce possess the largest ENMAX of 300 eV. As a consequence, setting the cutoff energy at 480 eV is large enough to get a satisfying result. As shown in Figure A1, when the cut off energy increases up to 500 eV, the computation time increase dramatically while the slab energy have no significant change. Likewise, we also test the convergence criteria of energy, convergence criteria of force and the distance threshold to fix atoms. The convergence test results are shown as figure A2.

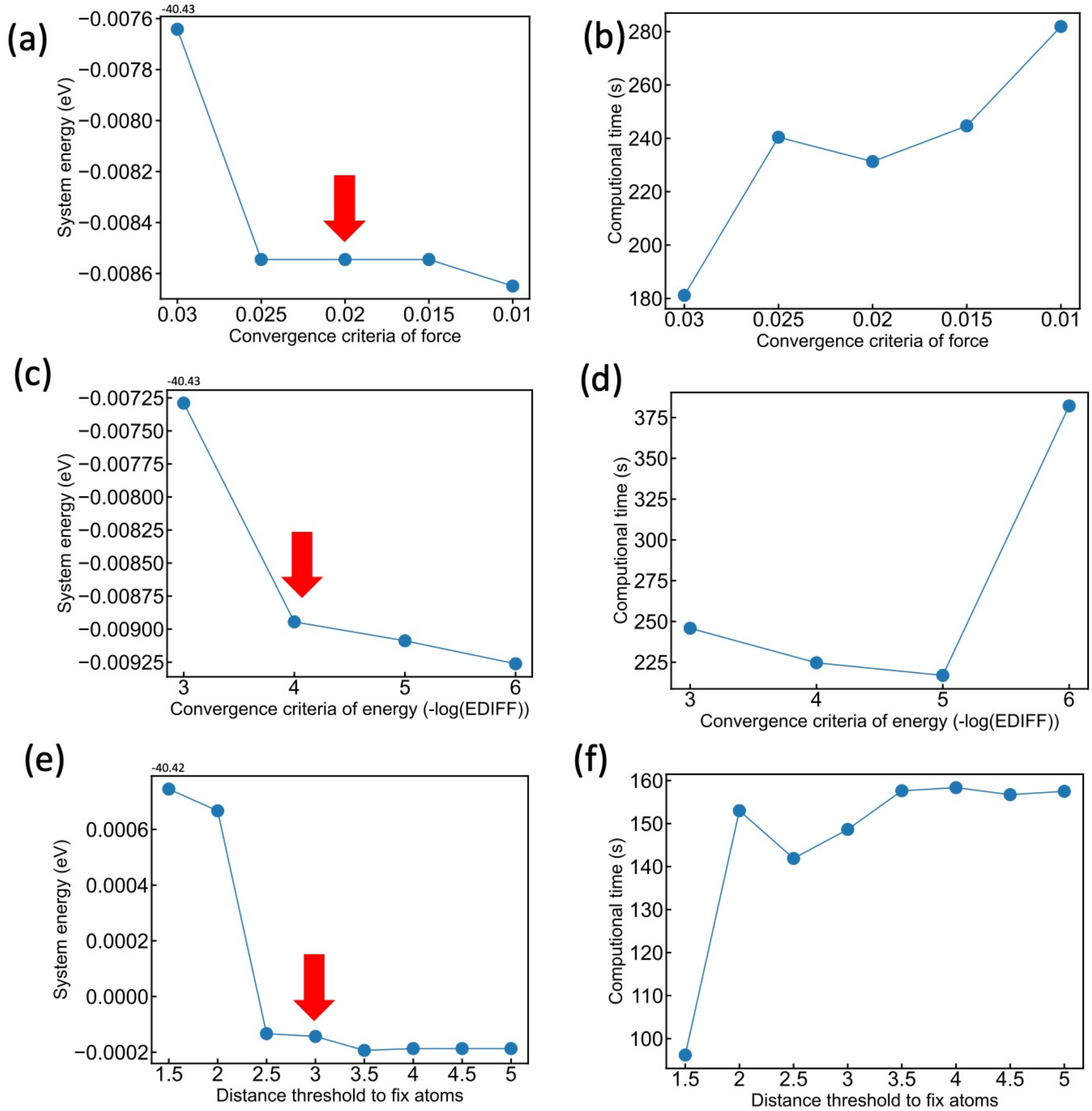

Figure A2. Convergence curves of MgSc slab model with respect to convergence criteria of energy, convergence criteria of force and the distance threshold to fix atoms.

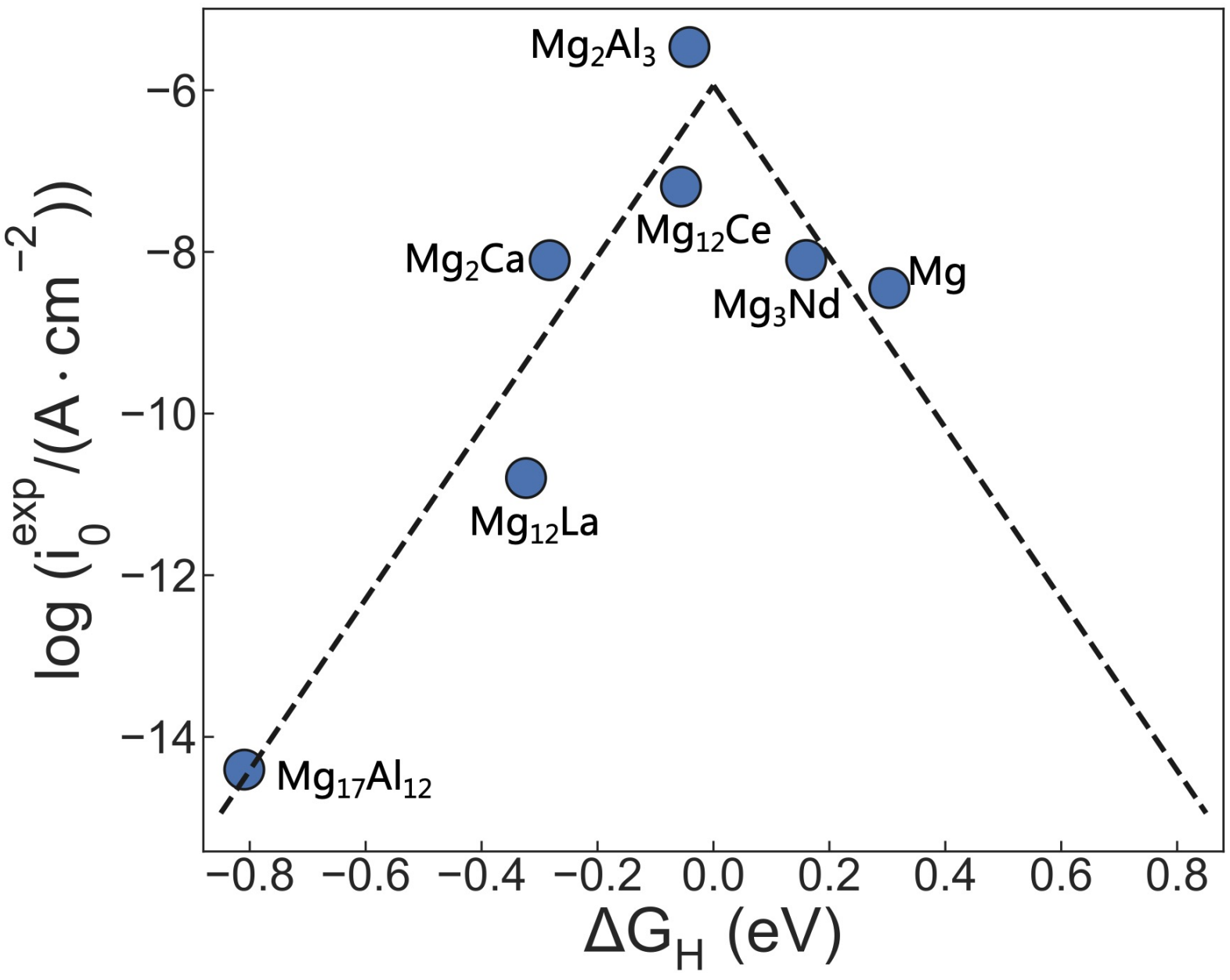

Figure A3. Volcano plot fitted by exchange current density of Mg and Mg intermetallics at pH=7, 0.1 mol/L Nacl.

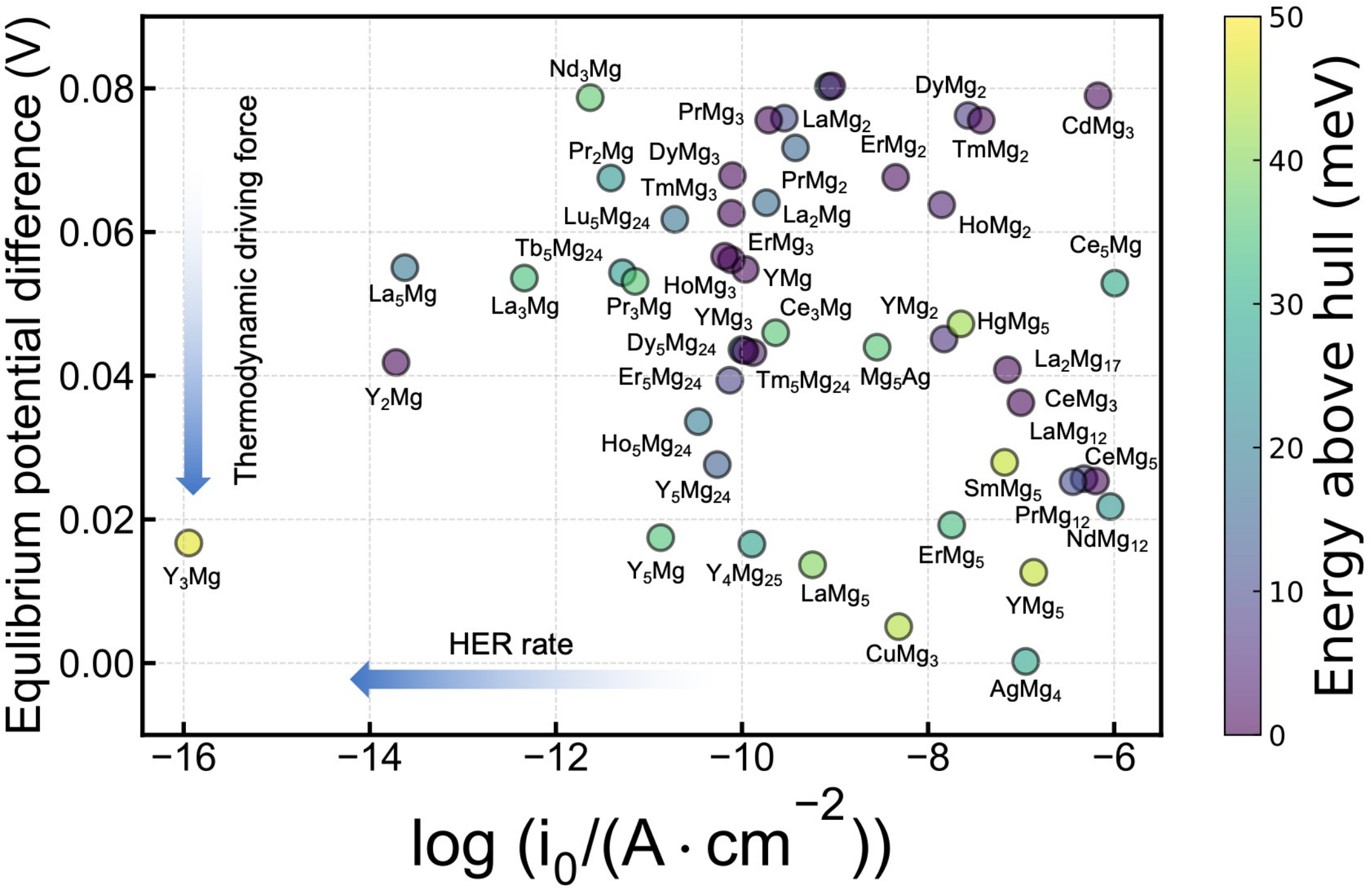

Figure A4. Comprehensive screening results containing phase stability, equilibrium potential difference and calculated exchange current density. The calculated exchange current density is calculated by the kinetic model, whose rate constant is fitted by Mg and Mg intermetallics at pH=7, 0.1 mol/L Nacl solution.

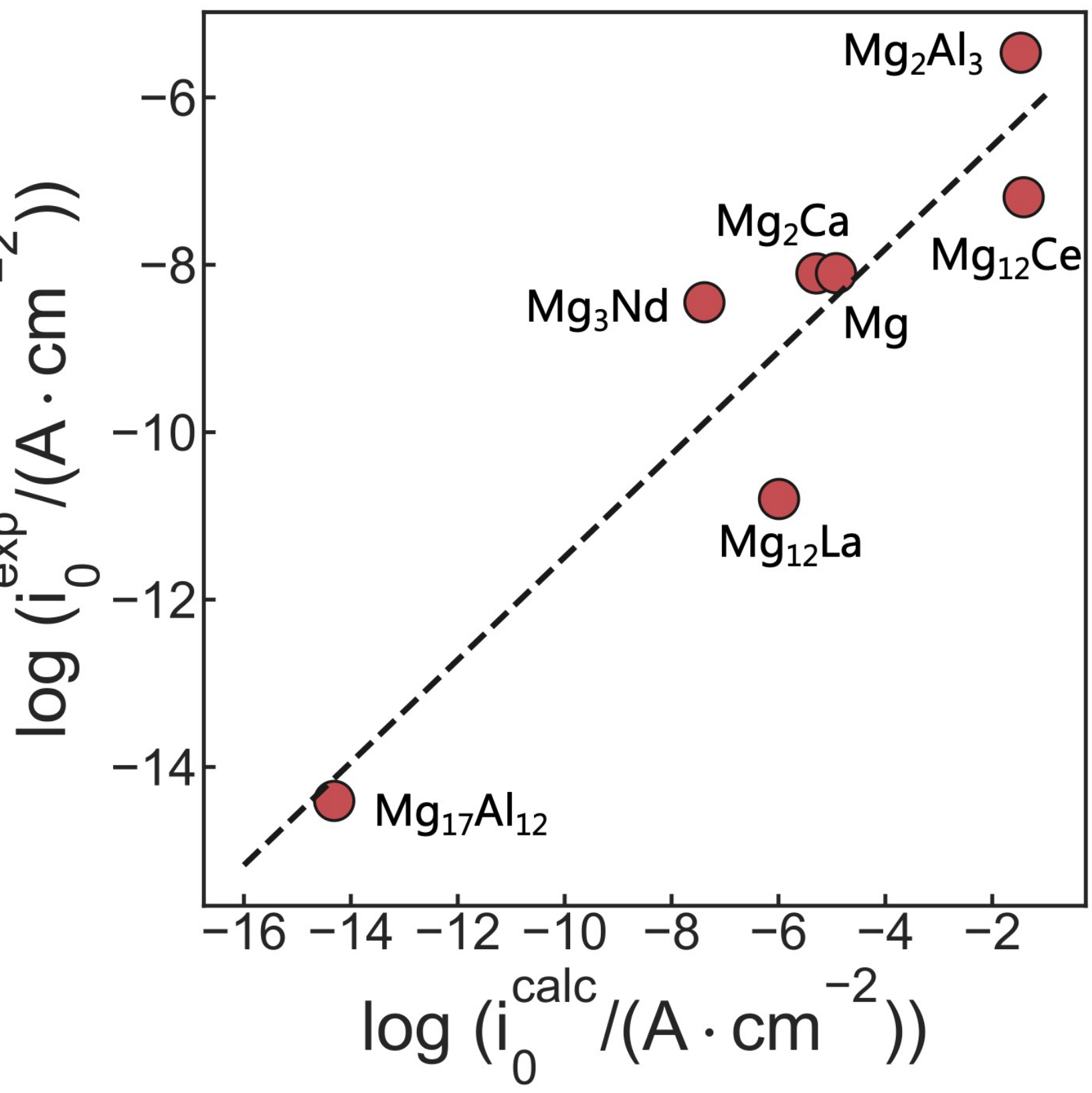

Figure A5. Comparison between experimental $i_0$ for some common binary Mg intermetallics at pH=7 and the $i_0$ calculated by the kinetic model, whose rate constant is fitted by experimental $i_0$ of some pure metals at pH=0.

Table A1. List of support vector regression (SVR) Hyperparameters and average Coefficient of determination ($R^2$) after 500 different random divisions of training (90 %) and test (10 %) sets. Other less sensitive hyperparameters are selected as default value in scikit-learn.

| **Algorithm** | **Ω** | **Features** | **Hyperparameter** | $R^2_{train}$ | $R^2_{test}$ | **RMSE** | **MAE** |
|---|---|---|---|---|---|---|---|
| KNN | 1 | $W_f$ | K = 4 | 0.592 | -0.040 | 0.239 | 0.169 |
| | 2 | $W_f$, WFIE | K = 2 | **0.927** | **0.667** | 0.128 | 0.097 |
| | 3 | $W_f$, WEN, RAM | K = 5 | 0.851 | 0.615 | 0.146 | 0.112 |
| | 4 | $W_f$, d, WEN, RAM | K = 3 | 0.891 | 0.605 | 0.144 | 0.108 |
| | 5 | $W_f$, $E_e$, EA, WFIE, RAM | K = 4 | 0.850 | 0.583 | 0.153 | 0.121 |
| SVR | 1 | $W_f$ | C=1, Gamma=1 | 0.400 | -0.121 | 0.256 | 0.195 |
| | 2 | $W_f$, WFIE | C=10, Gamma=1 | **0.941** | **0.677** | 0.133 | 0.102 |
| | 3 | $W_f$, WFIE, RAM | C=10, Gamma=1 | 0.969 | 0.656 | 0.137 | 0.107 |
| | 4 | $W_f$, WFIE, RAM, $E_e$ | C=1, Gamma=1 | 0.949 | 0.635 | 0.149 | 0.116 |
| | 5 | $W_f$, WFIE, RAM, $E_e$, WEN | C=10, Gamma=1 | 0.991 | 0.608 | 0.153 | 0.119 |

$c \in \{10^{-3}, 10^{-2}, 10^{-1}, 1, 10^{1}, 10^{2}, 10^{3}, 10^{4}, 10^{5}, 10^{6}, 10^{7}, 10^{8}\}$

gamma$\in \{10^{-8}, 10^{-7}, 10^{-6}, 10^{-5}, 10^{-4}, 10^{-3}, 10^{-3}, 10^{-2}, 10^{-1}, 1\}$

Table A2. $\Delta E_{ZPE}$ and $T\Delta S_H$ of HER for some common Mg intermetallics. In the unit of eV.

| System | Slab size | $\Delta E_{ZPE}$ | $T\Delta S_H(298K)$ | $\Delta E_{ZPE} - T\Delta S_H$ |
|---|---|---|---|---|
| Mg | 1*1 | -0.001 | -0.188 | 0.187 |
| Mg | 2*2 | 0.000 | -0.189 | 0.190 |
| Mg | 3*3 | 0.004 | -0.189 | 0.194 |
| $Mg_2Ca$ | 1*1 | 0.001 | -0.188 | 0.189 |
| $Mg_{24}Y_5$ | 1*1 | 0.003 | -0.188 | 0.191 |
| $Mg_3Nd$ | 1*1 | -0.046 | -0.181 | 0.135 |
| $Mg_2Si$ | 1*1 | -0.011 | -0.187 | 0.175 |
| $Mg_{12}Ce$ | 1*1 | 0.007 | -0.190 | 0.197 |
| $Mg_{17}Al_{12}$ | 1*1 | 0.010 | -0.184 | 0.193 |
| $MgZn_2$ | 1*1 | -0.021 | -0.195 | 0.174 |

Table A3. Calculated data of 50 stable binary Mg intermetallics with low equilibrium potential difference with respect to the Mg matrix. $\boldsymbol{\gamma}$ is the surface energy of intermetallics (mJ/m$^2$), $\boldsymbol{E_{ads}}$ is adsorption energy of hydrogen atom, $\Delta\boldsymbol{G_{H*}}$ is the free energy of the adsorbed state, $\boldsymbol{E_{Hull}}$ is the energy above hull of intermetallics at 0K, $\boldsymbol{E_e}$ is the equilibrium potential of intermetallics, $\boldsymbol{i_{0_pred}}$ is HER exchange current density on intermetallics fitted by volcano curve. A more detailed data including binary Mg intermetallic dissolution reaction and ML input features can be found at https://github.com/ywwang0/High-throughput-screens-corrosion-resistant-binary-magnesium-alloy.

| Intermetallic | Spacegroup | Miller index | $\boldsymbol{\gamma}$ | $\boldsymbol{E_{ads}}$ | $\Delta\boldsymbol{G_{H*}}$ | $\boldsymbol{E_{Hull}}$ | $\boldsymbol{E_e}$ | $i_0^{calc}$ |
|---|---|---|---|---|---|---|---|---|
| $Y_3Mg$ | Imm2 | 001 | 882.704 | -1.135 | -0.945 | 0.047 | -2.517 | -17.480 |
| $Y_2Mg$ | Cmcm | 010 | 828.986 | -0.924 | -0.734 | 0.000 | -2.492 | -13.881 |
| $La_5Mg$ | Cm | 010 | 787.801 | -0.915 | -0.725 | 0.019 | -2.479 | -13.728 |
| $La_3Mg$ | I4/mmm | 111 | 298.084 | -0.794 | -0.604 | 0.034 | -2.480 | -11.651 |
| $Nd_3Mg$ | Imm2 | 001 | 669.121 | -0.727 | -0.537 | 0.037 | -2.455 | -10.508 |
| $Pr_2Mg$ | I4/mmm | 111 | 781.326 | -0.706 | -0.516 | 0.025 | -2.466 | -10.150 |
| $Tb_5Mg_{24}$ | $I\bar{4}3m$ | 100 | 695.382 | -0.694 | -0.504 | 0.026 | -2.480 | -9.947 |
| $Pr_3Mg$ | Imm2 | 001 | 662.794 | -0.682 | -0.492 | 0.036 | -2.481 | -9.729 |
| $Y_5Mg$ | Cm | 001 | 893.781 | -0.656 | -0.466 | 0.035 | -2.517 | -9.283 |
| $Lu_5Mg_{24}$ | $I\bar{4}3m$ | 100 | 664.992 | -0.641 | -0.451 | 0.018 | -2.472 | -9.039 |
| $Ho_5Mg_{24}$ | $I\bar{4}3m$ | 100 | 691.399 | -0.618 | -0.428 | 0.020 | -2.500 | -8.631 |
| $Y_5Mg_{24}$ | $I\bar{4}3m$ | 100 | 777.576 | -0.598 | -0.408 | 0.014 | -2.506 | -8.301 |
| $Er_5Mg_{24}$ | $I\bar{4}3m$ | 100 | 770.225 | -0.586 | -0.396 | 0.009 | -2.495 | -8.084 |
| $Dy_5Mg_{24}$ | $I\bar{4}3m$ | 100 | 758.899 | -0.573 | -0.383 | 0.017 | -2.490 | -7.871 |
| YMg | Pm3m | 100 | 763.132 | -0.570 | -0.380 | 0.000 | -2.479 | -7.813 |
| $Y_4Mg_{25}$ | $R\bar{3}m$ | 110 | 713.127 | -0.563 | -0.373 | 0.028 | -2.517 | -7.700 |
| $Tm_5Mg_{24}$ | $I\bar{4}3m$ | 100 | 767.048 | -0.562 | -0.372 | 0.003 | -2.491 | -7.681 |
| $La_2Mg$ | C2/m | $10\bar{1}$ | 727.809 | -0.549 | -0.359 | 0.017 | -2.470 | -7.448 |
| $Ce_3Mg$ | Imm2 | 100 | 785.818 | -0.539 | -0.349 | 0.036 | -2.488 | -7.287 |
| $LaMg_2$ | $Fd\bar{3}m$ | 110 | 761.728 | -0.530 | -0.340 | 0.011 | -2.458 | -7.136 |
| $PrMg_2$ | $Fd\bar{3}m$ | 110 | 762.613 | -0.519 | -0.329 | 0.016 | -2.462 | -6.942 |
| $LaMg_5$ | $P\bar{6}2m$ | 100 | 366.701 | -0.502 | -0.312 | 0.040 | -2.520 | -6.647 |
| $Ho_3Mg$ | P4/mmm | 001 | 618.510 | -0.485 | -0.295 | 0.022 | -2.454 | -6.369 |
| $ErMg_2$ | $P6_3/mmc$ | 001 | 829.678 | -0.417 | -0.227 | 0.001 | -2.466 | -5.202 |
| $HoMg_2$ | $P6_3/mmc$ | 001 | 818.807 | -0.371 | -0.181 | 0.004 | -2.470 | -4.403 |
| $YMg_2$ | $P6_3/mmc$ | 001 | 808.704 | -0.368 | -0.178 | 0.006 | -2.489 | -4.361 |
| $ErMg_5$ | $P\bar{6}2m$ | 001 | 567.089 | -0.361 | -0.171 | 0.034 | -2.515 | -4.229 |
| $DyMg_2$ | $P6_3/mmc$ | 001 | 807.098 | -0.344 | -0.154 | 0.008 | -2.458 | -3.944 |
| $TmMg_2$ | $P6_3/mmc$ | 001 | 844.994 | -0.331 | -0.141 | 0.001 | -2.458 | -3.717 |
| $SmMg_5$ | $P\bar{6}2m$ | 001 | 467.775 | -0.307 | -0.117 | 0.045 | -2.506 | -3.310 |
| $AgMg_4$ | C2/m | 011 | 507.141 | -0.285 | -0.095 | 0.029 | -2.534 | -2.944 |
| $YMg_5$ | C2/m | 100 | 715.173 | -0.277 | -0.087 | 0.045 | -2.521 | -2.805 |
| $LaMg_{12}$ | I4/mmm | 101 | 683.089 | -0.226 | -0.036 | 0.009 | -2.508 | -1.930 |
| $CeMg_5$ | $P\bar{6}2m$ | 001 | 661.133 | -0.215 | -0.025 | 0.000 | -2.509 | -1.730 |
| $NdMg_{12}$ | I4/mmm | 101 | 683.901 | -0.200 | -0.010 | 0.025 | -2.512 | -1.474 |
| $Ce_5Mg$ | Cm | 001 | 785.393 | -0.195 | -0.005 | 0.030 | -2.481 | -1.396 |
| $GdMg_3$ | $Fm\bar{3}m$ | 111 | 716.459 | -0.168 | 0.022 | 0.000 | -2.455 | -1.698 |
| $PrMg_{12}$ | I4/mmm | 101 | 720.599 | -0.143 | 0.047 | 0.011 | -2.509 | -2.134 |
| $CeMg_3$ | $Fm\bar{3}m$ | 101 | 706.403 | -0.090 | 0.100 | 0.000 | -2.498 | -3.041 |
| $La_2Mg_{17}$ | $P6_3/mmc$ | 110 | 738.027 | -0.076 | 0.114 | 0.000 | -2.493 | -3.276 |
| $HgMg_5$ | Cm | 111 | 699.734 | -0.029 | 0.161 | 0.042 | -2.487 | -4.087 |
| $CuMg_3$ | P4/mmm | 001 | 352.454 | 0.034 | 0.224 | 0.044 | -2.529 | -5.169 |
| $AgMg_5$ | $P\bar{6}2m$ | 001 | 570.434 | 0.056 | 0.246 | 0.036 | -2.490 | -5.545 |
| $LaMg_3$ | I4/mmm | 101 | 772.018 | 0.102 | 0.292 | 0.000 | -2.454 | -6.334 |
| $PrMg_3$ | $Fm\bar{3}m$ | 111 | 753.694 | 0.166 | 0.356 | 0.000 | -2.458 | -7.433 |
| $YMg_3$ | $Fm\bar{3}m$ | 111 | 738.801 | 0.190 | 0.380 | 0.000 | -2.491 | -7.845 |
| $DyMg_3$ | $Fm\bar{3}m$ | 111 | 724.925 | 0.203 | 0.393 | 0.000 | -2.466 | -8.065 |
| $TmMg_3$ | $Fm\bar{3}m$ | 111 | 760.979 | 0.204 | 0.394 | 0.000 | -2.471 | -8.081 |
| $HoMg_3$ | $Fm\bar{3}m$ | 111 | 731.543 | 0.204 | 0.394 | 0.000 | -2.478 | -8.084 |
| $ErMg_3$ | $Fm\bar{3}m$ | 111 | 729.930 | 0.211 | 0.401 | 0.000 | -2.477 | -8.199 |

Table A4. The data of 5 new binary Mg intermetallics. Eads represents the adsorption energy of H. KNN and SVR represent the predicted adsorption energy of H via k-Nearest Neighbors and support vector regression algorithm. WF represents Work function (eV). WFIE represent the weighted first ionization energy (eV).

| Intermetallic | Source | Space group | Eads | KNN | SVR | WF | WFIE |
|---|---|---|---|---|---|---|---|
| LaMg | mp-1104 | $Pm\bar{3}m$ | -0.59 | -0.82 | -1.09 | 3.47 | 6.61 |
| Mg2Cu | mp-2481 | Fddd | -0.33 | -0.28 | -0.35 | 3.47 | 7.67 |
| Mg3Ag | mp-864952 | $P6_3/mmc$ | -0.13 | -0.11 | -0.05 | 3.73 | 7.63 |
| HoMg | mp-1199 | $Pm\bar{3}m$ | -0.62 | -0.64 | -0.63 | 3.60 | 6.83 |
| NdMg2 | mp-2389 | $Fd\bar{3}m$ | -0.36 | -0.37 | -0.41 | 3.77 | 6.94 |